\documentclass[review]{elsarticle}

\usepackage{lineno,hyperref}
\modulolinenumbers[1]
\usepackage{natbib,amsmath,amssymb,graphics,subfigure}
\setcitestyle{authoryear,round}
\journal{Journal of \LaTeX\ Templates}
\AtBeginDocument{\setlength\bibsep{1ex}}
\begin{document}

\begin{frontmatter}

\title{Traversable wormholes in $f(R)$ gravity with constant and variable redshift functions}
\author[mymainaddress]{Nisha Godani}
\ead{nishagodani.dei@gmail.com}

\author[mysecondaryaddress]{Gauranga C. Samanta\corref{mycorrespondingauthor}}
\cortext[mycorrespondingauthor]{Corresponding author}
\ead{gauranga81@gmail.com}

\address[mymainaddress]{Department of Mathematics, Institute of Applied Sciences and Humanities\\ GLA University, Mathura, Uttar Pradesh, India}
\address[mysecondaryaddress]{Department of Mathematics, BITS Pilani K K Birla Goa Campus, Goa, India}

\begin{abstract}
The present paper is aimed at the study of traversable wormholes in $f(R)$ gravity with a viable $f(R)$ function defined as
$f(R)=R-\mu R_c\Big(\frac{R}{R_c}\Big)^p$, where  $R$ is scalar curvature, $\mu$, $R_c$ and $p$ are constants with  $\mu, R_c>0$  and $0<p<1$ \citep{Amendola}. The metric of wormhole is dependent on shape function $b(r)$ and redshift function $\phi(r)$ which characterize its properties, so the shape function and redshift function play an important role in wormhole modeling. In this work, the wormhole solutions are determined for (i) $\phi(r)=\frac{1}{r}$ and
(ii) $\phi(r)=c$ (constant) with  $b(r)=\frac{r}{exp(r-r_0)}$ \citep{godani1}. Further, the regions respecting the energy conditions are investigated.
\end{abstract}


\end{frontmatter}


\section{Introduction}
Wormholes are hypothetical objects which have feature of connecting two distinct universes or two distinct points of the same universe. Initially, Flamm \citep{flamm} suggested the notion of wormhole. After that, Einstein and Rosen \citep{eins-ros} presented a geometrical construction of the same type which is known as Einstein-Rosen bridge. The concept of traversable wormholes was first introduced by Morris and Thorne \citep{morris1} for time travel. They proposed it as a tool for teaching general theory of relativity. They considered a spherically symmetric metric and marked the presence of exotic matter, the matter that does not respect energy conditions, at the throat to sustain the wormhole solutions with open throat. Indeed, the traversable wormhole solutions may not be obtained in general relativity, if the null energy condition is satisfied. This issue can be resolved by considering the systems where quantum effects compete with the classical ones \citep{Visser1, Gao, Maldacena, Caceres, Fu}. Alternatively, the presence of additional fields can also be considered as a source for the dissatisfaction of null energy condition (NEC) which has an association with various problematic instabilities \citep{Bronnikov1, Armendariz-Picon, Dubovsky, Bronnikov2, Gonzalez, Gonzaleza, Rubakov, Rubakov1, Evseev}.
The establishment of energy conditions in wormhole setting is an important issue which has been investigated in literature, for instance in dynamic and thin shell wormholes \citep{Kar, Arellano, Cataldo, Mehdizadeh}, by proposing new methods. Other than this, various researchers have tried to obtain wormhole solutions using the background  of modified theories of gravity. These solutions are developed in Kaluza-Klein gravity,  Born-Infeld theory, Brans-Dicke theory, Einstein-Gauss-Bonnet theory, Einstein-Cartan theory, scalar tensor theory etc. \citep{Agnese, Nandi, Dzhunushaliev, Bronnikov4, Richarte, Leon, Lobo1, Sushkov, Eiroa, Zangeneh, Bronnikov5, Shaikh, Bronnikov6, Mehdizadeh1, Bronnikov3}.

In modified theories of gravity, the stress energy tensor is replaced with effective stress energy tensor that contains curvature terms of higher order. The generalized theories of gravity are used to sort out the problem of exotic matter in wormholes, to construct viable cosmological models of our universe, to explain the singularities etc.   The $f(R)$ theory of gravity is one of the modified theories in which the geometrical part is modified by replacing scalar curvature $R$ in Lagrangian gravitation action by a general function $f(R)$. The field equations obtained with respect to this theory are highly complex and possess a larger set of solutions than general relativity. This theory is also simplified in \citep{Bertolami16} that provides a coupling between  matter and function $f(R)$ leading towards an extra force that may justify the current accelerating scenario of the universe \citep{Nojiri15, Bertolami46}.   Starobinsky \citep{Starobinsky99} presented the first model of inflation. Astashenok et al. \citep{Astashenok} explored the effects of various $f(R)$ gravity models on the evolution of
compact objects.
M. Sharif and Z. Yousaf \citep{Sharif1} studied charged adiabatic Lema\^{i}tre-Tolman-Bondi (LTB) gravitational collapse using $f(R)$ theory of gravity.
The mass function is obtained for cylindrical object using $f(R)$ gravity in comparison with the Misner–Sharp mass in spherical system \citep{Yousaf1}.
Yousaf et al. \citep{Yousaf2} considered Lema\^{i}tre-Tolman-Bondi dynamical model in the form of compact object and investigated its evolution in the context of both tilted and non-tilted observers using the background of $f(R)$ gravity. Further, the compact objects are studied using modified $f(R)$ gravity in  \citep{Hwang, Sharif2, Artyom, Nojiri2014, Sharif3}.
  Many other cosmological models are studied from different aspects in the context of $f(R)$ gravity \citep{Capozziello91, Bombacigno21, Sbis, Chen, Elizalde26, Elizalde25, Astashenok1, Miranda, Nascimento, Odintsov20, Odintsov1, Nojiri56, Parth}.

 In literature, various other modified theories of gravity are also introduced which include $f(R,T)$ gravity, where $T$ is the trace of stress energy tensor, $f(G)$ gravity, where $G$ stands for the Gauss Bonnet invariant, and $f(R, T, R_{\mu\nu}, T^{\mu\nu})$ gravity. We discuss some literature in the scenario of gravitational collapse using these modified theories of gravity.
 Yousaf et al. \citep{Yousaf3} considered spherical geometry coupled with heat and radiation and explored its behavior using $f(R,T)$ theory of gravity. Bamba et al.  \citep{Bamba1} taken into account the modified $f(G)$ gravity to investigate the energy conditions for flat Friedmann-Lema\^{i}tre-Robertson-Walker metric with perfect fluid. They also found the viability bounds of the model by null and weak energy conditions.
 Yousaf et al. \citep{Yousaf4} studied the dynamical instability of spherical anisotropic sources in the context of $f(R, T, R_{\mu\nu}, T^{\mu\nu})$ gravity.
 Yousaf et al. \citep{Yousaf5} examined the role of electromagnetic field on spherical gravastar models in $f(R,T)$  gravity.
 Yousaf  \citep{Yousaf6} considered irrotational cylindrically symmetric geometry and studied  the role of electromagnetic field on the viable matching conditions between exterior and interior geometries using the framework of $f(R,T)$ gravity. Bhatti et al. \citep{Bhatti3} obtained the constraints on physical quantities for the stability of  celestial self-gravitating configurations with anisotropic environment.

The $f(R)$ theory of gravity has been extensively used in the investigation of wormhole solutions.  The static wormholes using the non-commutative geometry are developed \citep{Rahaman, Jamil}.  The junction conditions in  $f(R)$ gravity are  applied to build pure double layer bubbles and thin shell \citep{Eiroa1, Eiroa2, Eiroa3, Eiroa4}. The cosmological development of wormhole solutions is explored in \citep{Bhattacharya}. Dynamical wormholes without need of exotic matter and asymptotically tending to FLRW universe are obtained in \citep{Bahamonde}. Lorentzian wormhole solutions are analyzed with viable $f(R)$  model in \citep{Pavlovic}.
Traversable wormhole solutions are constructed in $f(R)$ gravity and higher order curvature terms are found to be responsible for the dissatisfaction of NEC \citep{Lobo12}. Taking constant shape and redshift functions, the energy conditions for wormhole geometries are examined in \citep{saiedi}. However, with novel and variable shape function and constant redshift function, these are examined in \citep{peter, godani, godani1}. Further, the efforts are put to obtain the wormhole solutions with less amount of exotic matter using viable $f(R)$ gravity models \citep{Samanta19, Godani19, Samantaepjc}. Wormholes are also studied form different points of view in \citep{Kar2, Wang, Roman, Poisson, Barcelo, Gonz, Visser2, Gonz1, Sushkov1, Lobo, Lobo2, Bohmer1, Dotti, Forghani, Heydarzade, Moradpour}.

Other than the above attempts, Pavlovic and Sossich \citep{Pavlovic} explored wormhole solutions using four viable $f(R)$ gravity models which include the MJWQ model \citep{Miranda}, the exponential gravity model \citep{Cognola, Elizalde}, the Tsujikawa model \citep{Tsujikawa, Felice} and the Starobinsky model \citep{Tsujikawa, Starobinsky, Amendola1, Amendola3}. They defined the redshift function as $\phi(r)=\ln(\frac{r_0}{r}+1)$ and consequently, found the wormhole solutions filled with non-exotic matter.
Bhatti et al. \citep{Bhatti1} studied thin-shell wormholes for the charged black string using $f(R)$ theory of gravity for logarithmic and exponential form of $f(R)$  models. Bhatti et al. \citep{Bhatti2} investigated wormhole solutions using $f(R,T)$ gravity with model $f(R,T)=f(R)+\lambda T$, where $T$ is the trace of stress energy tensor. They considered $f(R)=R+\alpha R(\exp(\frac{R}{\gamma})-1)$ model with constant redshift function for three types of matter configurations.

Now, from these preceding studies carried out in the past, it can be observed that the researchers have been attempting to develop wormhole solutions with or without exotic matter for different choices of $f(R)$ model, redshift and shape functions. Since the normal matter obeys the energy conditions, therefore the interest seems in the  development of wormholes without exotic matter. Like various researchers, in this paper, we have also adopted the direction of avoiding the exotic matter for particular choices of functions required for the model. Now, we are going to discuss the motivation for the choice of model functions considered in this work.
Samanta and Godani \citep{Samanta19} used the viable $f(R)=R-\mu R_c\Big(\frac{R}{R_c}\Big)^p$ model   with shape function $b(r)=\frac{r_0\log(r+1)}{\log(r_0+1)}$ and constant redshift function. They obtained wormhole solutions without need of exotic matter, i.e they found the satisfaction of energy conditions. The energy conditions may or may not be satisfied for this $f(R)$ function with different shape and redshift functions. To explore this possibility, we have adopted the same $f(R)$ function in the present paper. Further,
Anchordoqui et al. \citep{Anchordoqui} used variable redshift function as $\phi(r)=-\frac{\alpha}{r}$ with $\alpha>0$ and determined analytical wormhole solutions. Sarkar et al. \citep{Sarkar} assumed  $\phi(r)=\frac{\alpha}{r}$, where $\alpha$ is a constant, and obtained wormhole solutions in $\kappa(R,T)$ gravity and obtained wormhole solutions
filled with exotic matter. Further, Rahaman et al. \citep{Rahaman} used $\phi(r)= \frac{\alpha}{r}$ to obtain the generating functions comprising
the wormhole geometry. Further for $\phi(r)= \frac{\alpha}{r}$, they obtained shape
function with a specific form of generating function. These studies motivate us to consider both constant and variable redshift function $\phi(r)=\frac{\alpha}{r}$  for the exploration of wormhole solution using a specific $f(R)$ gravity model. In particular, we have considered $\phi(r)=\frac{1}{r}$ and $f(R)=R-\mu R_c\Big(\frac{R}{R_c}\Big)^p$, where  $\mu$, $R_c$ and $p$ are constants with  $\mu, R_c>0$  and $0<p<1$ \citep{Amendola} in this study. Furthermore, the reason for the choice of shape function also depends on previous  studies. Godani et al. \citep{godani1} proposed the shape function as $b(r)=\frac{r}{\exp(r-r_0)}$ and explored energy conditions in both $f(R)$ and $f(R,T)$ theories of gravity using constant redshift function. They found the dissatisfaction of energy conditions in case of $f(R)$ gravity. Recently, Godani and Samanta \citep{ng2020} studied traversable wormholes in $f(R)=R+\alpha R^n$ gravity, where $\alpha$ and $n$ are arbitrary constants, with $\phi(r)=\ln(\frac{r_0}{r}+1)$ and $b(r)=\frac{r_0}{\exp(r-r_0)}$. They  obtained the wormhole solutions by avoiding the exotic matter for wormholes having radius of the throat greater than 1.6 for any $n<0$. This shows a change in the nature of energy conditions with the change in $\phi(r)$ and $f(R)$ functions.   To explore the possibility of having the traversable wormhole solutions completely filled with non-exotic matter, we have considered $b(r)=\frac{r_0}{\exp(r-r_0)}$.
Thus, the purpose of the present article is to obtain the wormhole solutions in modified $f(R)$ theory of gravity with a viable model $f(R)=R-\mu R_c\Big(\frac{R}{R_c}\Big)^p$, where  $R$ is Ricci scalar curvature, $\mu$, $R_c$ and $p$ are constants with  $\mu, R_c>0$  and $0<p<1$ \citep{Amendola}. The shape function  $b(r)=\frac{r}{exp(r-r_0)}$ \citep{godani1} is taken with both constant and variable redshift functions to detect the validity of energy conditions.

\section{Field Equations \& Wormhole Geometry}
The static and spherically symmetric  metric defining the wormhole structure  is
\begin{equation}\label{metric}
ds^2=-e^{2\Phi(r)}dt^2+\frac{dr^2}{1-b(r)/r} + r^2(d\theta^2+\sin^2\theta d\phi^2).
\end{equation}
The  function $\Phi (r)$ determines the gravitational redshift, hence it is called redshift function.
The wormhole solutions must satisfy Einstein's field equations and must possess a throat that joins two regions of universe which are asymptotically flat. For a traversable wormhole, event horizon should not be present and the effect of tidal gravitational forces should be very small on a traveler.

The functions $\Phi(r)$ and $b(r)$ are the functions of  radial coordinate $r$, which is a non-decreasing function. Its minimum value is $r_0$, radius of the throat, and maximum value is $+\infty$.  The function $b(r)$ determines the shape of wormhole, hence it is called as shape function. The existence of wormhole solutions demands the satisfaction of the following conditions:
(i) $b(r_0)=r_0$, (ii) $\frac{b(r)-b'(r)r}{b^2}>0$, (iii) $b'(r_0)-1\leq 0$, (iv) $\frac{b(r)}{r}<1$ for $r>r_0$ and (v) $\frac{b(r)}{r}\rightarrow 0$ as $r\rightarrow\infty$. For simplicity, the redshift function is assumed as a constant.

Morris \& Thorne \citep{morris1} introduced traversable wormholes in the framework of Einstein's general theory of relativity. The $f(R)$  theory of gravity is a generalization of  Einstein's theory of relativity which  replaces the  gravitational action $R$ with a general function $f(R)$ of $R$. Thus, the  gravitational action for $f(R)$ theory of gravity is defined as
\begin{equation}\label{action}
S_G=\dfrac{1}{2k}\int[f(R) + L_m]\sqrt{-g}d^4x,
\end{equation}
where $k=8\pi G$, $L_m$ and $g$ stand for the  matter Lagrangian density and  the  determinant of the metric $g_{\mu\nu}$ respectively. For simplicity, $k$ is taken as unity.\\

Variation of Eq.(\ref{action}) with respect to the metric $g_{\mu\nu}$ gives the field equations as
\begin{equation}\label{fe}
FR_{\mu\nu} -\dfrac{1}{2}fg_{\mu\nu}-\triangledown_\mu\triangledown_\nu F+\square Fg_{\mu\nu}= T_{\mu\nu}^m,
\end{equation}	
where $R_{\mu\nu}$ and $R$ denote Ricci tensor and  scalar curvature respectively and $F=\frac{df}{dR}$. The contraction of \ref{fe}, gives
\begin{equation}\label{trace}
FR-2f+3\square F=T,
\end{equation}
where $T=T^{\mu}_{\mu}$ is the trace of the stress energy tensor.

From Eqs. \ref{fe} \& \ref{trace}, the effective field equations are obtained as
\begin{equation}
G_{\mu\nu}\equiv R_{\mu\nu}-\frac{1}{2}Rg_{\mu\nu}=T_{\mu\nu}^{eff},
\end{equation}
where $T_{\mu\nu}^{eff}=T_{\mu\nu}^{c}+T_{\mu\nu}^{m}/F$ and $T_{\mu\nu}^{c}=\frac{1}{F}[\triangledown_\mu\triangledown_\nu F-\frac{1}{4}g_{\mu\nu}(FR+\square F+T)]$.
The energy momentum tensor for the matter source of the wormholes is $T_{\mu\nu}=\frac{\partial L_m}{\partial g^{\mu\nu}}$, which is defined as
\begin{equation}
T_{\mu\nu} = (\rho + p_t)u_\mu u_\nu - p_tg_{\mu\nu}+(p_r-p_t)X_\mu X_\nu,
\end{equation}	
such that
\begin{equation}
u^{\mu}u_\mu=-1 \mbox{ and } X^{\mu}X_\mu=1,
\end{equation}

where $\rho$,  $p_t$ and $p_r$  stand for the energy density, tangential pressure and radial pressure respectively.

Einstein's field equations for the metric \ref{metric} in  $f(R)$ gravity are obtained as:
\citep{lobo}:
\begin{equation}\label{6}
\rho=\frac{Fb'(r)}{r^2}-\Bigg(1-\frac{b(r)}{r}\Bigg)F'\phi'(r)-H
\end{equation}
\begin{equation}\label{7}
p_r=-\frac{b(r)F}{r^3}+2\Bigg(1-\frac{b(r)}{r}\Bigg)\frac{\phi^{'}(r)F}{r}-\Bigg(1-\frac{b(r)}{r}\Bigg)\Bigg[F''+\frac{F'(rb'(r)-b(r))}{2r^2\Big(1-\frac{b(r)}{r}\Big)}\Bigg]+H
\end{equation}
\begin{eqnarray}\label{8}
p_t&=&\frac{F(b(r)-rb'(r))}{2r^3}-\frac{F'}{r}\Bigg(1-\frac{b(r)}{r}\Bigg)+F\Bigg(1-\frac{b(r)}{r}\Bigg)\Bigg(\phi^{''}(r)\nonumber\\
&-&\frac{(rb'(r)-b(r))\phi^{'}(r)}{2r(r-b)}+\phi^{'2}(r)+\frac{\phi^{'}(r)}{r}\Bigg)+H,
\end{eqnarray}
where $H=\frac{1}{4}(FR+\square F+T)$ and prime upon a function denotes the derivative of that function with respect to  radial coordinate $r$.

\section{Wormhole Solutions}

Amendola et al. \citep{Amendola} explored cosmological viability of $f(R)$  models and examined their cosmological nature.  They derived autonomous equations for arbitrary $f(R)$ models, determined all fixed points for such system and examined the stability of these points to study the cosmological evolution. They drew $m(r)$ curves in the $rm$-plane, where $m=\frac{R\frac{dF}{dR}}{F}$, $r=-\frac{RF}{f}$ \& $F=\frac{df}{dR}$, and classified $f(R)$ models into four classes and tested cosmological viability. The model with
\begin{equation}\label{model}
f(R)=R-\mu R_c\Big(\frac{R}{R_c}\Big)^p,
\end{equation}
where $\mu, R_c>0$  and $0<p<1$, was obtained to belong to Class II and meeting all the conditions which consequently provide an acceptable cosmology.

A viable $f(R)$ dark energy model should fulfill the following conditions: (i) $F>0$ for $R\geq R_0$, where $R_0$ is the Ricci scalar at the present epoch, if the final attractor is a de Sitter point with the Ricci scalar $R_1$, then $F> 0$ for $R \geq R_1$; (ii) $\frac{dF}{dR}>0$ for $R\geq R_0$; (iii) $F\rightarrow R-\Lambda$ for $R>>R_0$ and (iv) $0<m<1$ at $r=-2$.
Model (\ref{model}) satisfies these four conditions. To satisfy condition (iii), the power $p$ in model (\ref{model}) should be close to zero.
The experimental bound on model  (\ref{model}) is obtained as $\frac{p}{2-p}\Big(\frac{R_1}{\rho_B}\Big)^{1-p}<1.5\times 10^{-15}$, where  $\rho_B$ is the density outside the body. For $R_1=10^{-29}$ g/cm$^3$ \& $\rho_B=10^{-24}$ g/cm$^3$, the constraint on $p$ is $p<3\times 10^{-10}$ that represents a very small change from $\Lambda$CDM model \citep{Dark}. These characteristics have inspired to consider the model \ref{model} in the present study.
Various viable $f(R)$ models, compatible with observations, are defined and studied in literature.
In this paper, the model \eqref{model} is considered to obtain the traversable  wormhole solutions  with the shape function $b(r)=\frac{r}{exp(r-r_0)}$ \citep{godani1}.
We are interested to determine these solutions with both constant and variable redshift functions $\phi(r)$. We have considered the variable redshift function $\phi(r)=\frac{1}{r}$. Further, we have solved the field equations and calculated the expressions for various combinations of the energy density ($\rho$) and pressures ($p_r$ and $p_t$) in the following two cases:\\

\noindent
\textbf{ Case 1:} $\phi(r)=c$ (constant)\\
The following expressions are obtained for $r\neq 1$ from the field equations \eqref{6} to \eqref{8} by substituting constant redshift function $\phi(r)=c$:
\begin{eqnarray}
\rho&=&\frac{1}{4 (r-1)^3}\Bigg[R_c \mu 2^{p+1} p \left(p^2 \left(r^2-2\right)^2-p \left(2 r^4+2 r^3-11 r^2+10\right)+r^4+2 r^3-7 r^2+6\right) e^{r-r_0}\nonumber\\
&\times& \left(\frac{(r-1) e^{r_0-r}}{R_c r^2}\right)^p+R_c \mu 2^p \left(-2 p^3 \left(r^2-2\right)^2+p^2 \left(5 r^4+3 r^3-24 r^2+2 r+20\right)\right.\nonumber\\
&+&\left.p \left(-3 r^4+r^3+4 r^2+10 r-16\right)-2 (r-1)^3\right) \left(\frac{(r-1) e^{r_0-r}}{R_c r^2}\right)^p-\frac{4 (r-1)^4 e^{r_0-r}}{r^2}\Bigg]
\end{eqnarray}

\begin{eqnarray}
p_r&=&-\frac{1}{(r-1)^2}\Bigg[R_c \mu 2^{p-1} \left(2 p^2 \left(r^2-2\right)-p \left(r^2+3 r-6\right)-(r-1)^2\right) \left(\frac{(r-1) e^{r_0-r}}{R_c r^2}\right)^p\Bigg]\nonumber\\
&+&\frac{1}{(r-1) r^2}\Bigg[\mu 2^p (p-1) p \left(r^2-2\right) \left(\frac{(r-1) e^{r_0-r}}{R_c r^2}\right)^{p-1}\Bigg]-\frac{e^{r_0-r}}{r^2}
\end{eqnarray}

\begin{eqnarray}
p_t&=&\frac{1}{4} \Bigg[-\frac{1}{(r-1)^2 r^2}\Bigg(\mu 2^{p+1} p \left(p^2 \left(r^2-2\right)^2-p \left(2 r^4+r^3-10 r^2+2 r+8\right)+r^4+r^3-6 r^2\right.\nonumber\\
&+&\left.2 r+4\right) \left(\frac{(r-1) e^{r_0-r}}{R_c r^2}\right)^{p-1}\Bigg)+\frac{1}{(r-1)^3}\Bigg(R_c \mu 2^p \left(2 p^3 \left(r^2-2\right)^2-p^2 \left(5 r^4+r^3-22 r^2\right.\right.\nonumber\\
&+&\left.\left.6 r+16\right)+p \left(3 r^4-2 r^3-6 r^2-r+10\right)+2 (r-1)^3\right) \left(\frac{(r-1) e^{r_0-r}}{R_c r^2}\right)^p\Bigg)+\frac{2 e^{r_0-r}}{r}\Bigg]
\end{eqnarray}

\begin{eqnarray}
\rho+p_r&=&-\frac{1}{(r-1)^2}\Bigg[R_c \mu 2^{p-1} \left(2 p^2 \left(r^2-2\right)-p \left(r^2+3 r-6\right)-(r-1)^2\right) \left(\frac{(r-1) e^{r_0-r}}{R_c r^2}\right)^p\Bigg]\nonumber\\
&+&\frac{1}{4 (r-1)^3}\Bigg[R_c \mu 2^{p+1} p \left(p^2 \left(r^2-2\right)^2-p \left(2 r^4+2 r^3-11 r^2+10\right)+r^4+2 r^3-7 r^2+6\right) \nonumber\\
&\times&e^{r-r_0} \left(\frac{(r-1) e^{r_0-r}}{R_c r^2}\right)^p+R_c \mu 2^p \left(-2 p^3 \left(r^2-2\right)^2+p^2 \left(5 r^4+3 r^3-24 r^2+2 r+20\right)\right.\nonumber\\
&+&\left.p \left(-3 r^4+r^3+4 r^2+10 r-16\right)-2 (r-1)^3\right) \left(\frac{(r-1) e^{r_0-r}}{R_c r^2}\right)^p-\frac{4 (r-1)^4 e^{r_0-r}}{r^2}\Bigg]\nonumber\\
&+&\frac{\mu 2^p (p-1) p \left(r^2-2\right) \left(\frac{(r-1) e^{r_0-r}}{R_c r^2}\right)^{p-1}}{(r-1) r^2}-\frac{e^{r_0-r}}{r^2}
\end{eqnarray}

\begin{eqnarray}
\rho+p_t&=&\frac{1}{4} \Bigg[-\frac{1}{(r-1) r^2}\Bigg(\mu 2^{p+1} (p-1) p \left(r^2-2\right) \left(\frac{(r-1) e^{r_0-r}}{R_c r^2}\right)^{p-1}\Bigg)+\frac{1}{(r-1)^2} \nonumber\\
&\times&\Bigg(R_c \mu 2^p p \left(2 p \left(r^2-2\right)-r^2-3 r+6\right) \left(\frac{(r-1) e^{r_0-r}}{R_c r^2}\right)^p\Bigg)-\frac{2 (r-2) e^{r_0-r}}{r^2}\Bigg]
\end{eqnarray}

\begin{eqnarray}
\rho+p_r+2p_t&=&\frac{1}{(r-1)^3}\Bigg[R_c \mu 2^{p-2} (p-1) e^{-r_0} \left(e^{r_0} \left(2 p^2 \left(r^2-2\right)^2-p \left(3 r^4+3 r^3-16 r^2+2 r+12\right)\right.\right.\nonumber\\
&-&\left.\left.4 (r-1)^3\right)-2 p e^r \left(p \left(r^2-2\right)^2-r^4-2 r^3+7 r^2-6\right)\right) \left(\frac{(r-1) e^{r_0-r}}{R_c r^2}\right)^p\Bigg]
\end{eqnarray}
\begin{eqnarray}
\rho-|p_r|&=&\frac{1}{4 (r-1)^3}\Bigg[R_c \mu 2^{p+1} p \left(p^2 \left(r^2-2\right)^2-p \left(2 r^4+2 r^3-11 r^2+10\right)+r^4+2 r^3-7 r^2+6\right) e^{r-r_0}\nonumber\\
&\times& \left(\frac{(r-1) e^{r_0-r}}{R_c r^2}\right)^p+R_c \mu 2^p \left(-2 p^3 \left(r^2-2\right)^2+p^2 \left(5 r^4+3 r^3-24 r^2+2 r+20\right)\right.\nonumber\\
&+&\left.p \left(-3 r^4+r^3+4 r^2+10 r-16\right)-2 (r-1)^3\right) \left(\frac{(r-1) e^{r_0-r}}{R_c r^2}\right)^p-\frac{4 (r-1)^4 e^{r_0-r}}{r^2}\Bigg]\nonumber\\
&-&\Bigg|
-\frac{1}{(r-1)^2}\Bigg[R_c \mu 2^{p-1} \left(2 p^2 \left(r^2-2\right)-p \left(r^2+3 r-6\right)-(r-1)^2\right) \left(\frac{(r-1) e^{r_0-r}}{R_c r^2}\right)^p\Bigg]\nonumber\\
&+&\frac{1}{(r-1) r^2}\Bigg[\mu 2^p (p-1) p \left(r^2-2\right) \left(\frac{(r-1) e^{r_0-r}}{R_c r^2}\right)^{p-1}\Bigg]-\frac{e^{r_0-r}}{r^2}
\Bigg|
\end{eqnarray}

\begin{eqnarray}
\rho-|p_t|&=&\frac{1}{4 (r-1)^3}\Bigg[R_c \mu 2^{p+1} p \left(p^2 \left(r^2-2\right)^2-p \left(2 r^4+2 r^3-11 r^2+10\right)+r^4+2 r^3-7 r^2+6\right) e^{r-r_0}\nonumber\\
&\times& \left(\frac{(r-1) e^{r_0-r}}{R_c r^2}\right)^p+R_c \mu 2^p \left(-2 p^3 \left(r^2-2\right)^2+p^2 \left(5 r^4+3 r^3-24 r^2+2 r+20\right)\right.\nonumber\\
&+&\left.p \left(-3 r^4+r^3+4 r^2+10 r-16\right)-2 (r-1)^3\right) \left(\frac{(r-1) e^{r_0-r}}{R_c r^2}\right)^p-\frac{4 (r-1)^4 e^{r_0-r}}{r^2}\Bigg]\nonumber\\
&-&\Bigg|\frac{1}{4} \Bigg[-\frac{1}{(r-1)^2 r^2}\Bigg(\mu 2^{p+1} p \left(p^2 \left(r^2-2\right)^2-p \left(2 r^4+r^3-10 r^2+2 r+8\right)+r^4+r^3-6 r^2\right.\nonumber\\
&+&\left.2 r+4\right) \left(\frac{(r-1) e^{r_0-r}}{R_c r^2}\right)^{p-1}\Bigg)+\frac{1}{(r-1)^3}\Bigg(R_c \mu 2^p \left(2 p^3 \left(r^2-2\right)^2-p^2 \left(5 r^4+r^3-22 r^2\right.\right.\nonumber\\
&+&\left.\left.6 r+16\right)+p \left(3 r^4-2 r^3-6 r^2-r+10\right)+2 (r-1)^3\right) \left(\frac{(r-1) e^{r_0-r}}{R_c r^2}\right)^p\Bigg)+\frac{2 e^{r_0-r}}{r}\Bigg]\Bigg|
\end{eqnarray}

\begin{eqnarray}
p_t-p_r&=&\frac{1}{(r-1)^2}\Bigg[R_c \mu 2^{p-1} \left(2 p^2 \left(r^2-2\right)-p \left(r^2+3 r-6\right)-(r-1)^2\right) \left(\frac{(r-1) e^{r_0-r}}{R_c r^2}\right)^p\Bigg]\nonumber\\
&+&\frac{1}{4} \left(-\frac{1}{(r-1)^2 r^2}\Bigg[\mu 2^{p+1} p \left(p^2 \left(r^2-2\right)^2-p \left(2 r^4+r^3-10 r^2+2 r+8\right)+r^4+r^3\right.\right.\nonumber\\
&-&\left.\left.6 r^2+2 r+4\right) \left(\frac{(r-1) e^{r_0-r}}{R_c r^2}\right)^{p-1}\Bigg]+\frac{1}{(r-1)^3}\Bigg[R_c \mu 2^p \left(2 p^3 \left(r^2-2\right)^2-p^2 \left(5 r^4+r^3\right.\right.\right.\nonumber\\
&-&\left.\left.\left.22 r^2+6 r+16\right)+p \left(3 r^4-2 r^3-6 r^2-r+10\right)+2 (r-1)^3\right) \left(\frac{(r-1) e^{r_0-r}}{R_c r^2}\right)^p\Bigg]\right.\nonumber\\
&+&\left.\frac{2 e^{r_0-r}}{r}\right)-\frac{\mu 2^p (p-1) p \left(r^2-2\right) \left(\frac{(r-1) e^{r_0-r}}{R_c r^2}\right)^{p-1}}{(r-1) r^2}+\frac{e^{r_0-r}}{r^2}
\end{eqnarray}

\begin{eqnarray}
\frac{p_r}{\rho}&=&-\frac{1}{(r-1)^2}\Bigg[R_c \mu 2^{p-1} \left(2 p^2 \left(r^2-2\right)-p \left(r^2+3 r-6\right)-(r-1)^2\right) \left(\frac{(r-1) e^{r_0-r}}{R_c r^2}\right)^p\Bigg]\nonumber\\
&+&\frac{1}{(r-1) r^2}\Bigg[\mu 2^p (p-1) p \left(r^2-2\right) \left(\frac{(r-1) e^{r_0-r}}{R_c r^2}\right)^{p-1}\Bigg]-\frac{e^{r_0-r}}{r^2}\nonumber\\
&\div&\Bigg[ \frac{1}{4 (r-1)^3}\Bigg[R_c \mu 2^{p+1} p \left(p^2 \left(r^2-2\right)^2-p \left(2 r^4+2 r^3-11 r^2+10\right)+r^4+2 r^3-7 r^2+6\right) e^{r-r_0}\nonumber\\
&\times& \left(\frac{(r-1) e^{r_0-r}}{R_c r^2}\right)^p+R_c \mu 2^p \left(-2 p^3 \left(r^2-2\right)^2+p^2 \left(5 r^4+3 r^3-24 r^2+2 r+20\right)\right.\nonumber\\
&+&\left.p \left(-3 r^4+r^3+4 r^2+10 r-16\right)-2 (r-1)^3\right) \left(\frac{(r-1) e^{r_0-r}}{R_c r^2}\right)^p-\frac{4 (r-1)^4 e^{r_0-r}}{r^2}\Bigg]\Bigg]
\end{eqnarray}
\noindent
\textbf{ Case 2:} $\phi(r)=\frac{1}{r}$  \\
The following values are obtained for $r\neq 1$ from the field equations \eqref{6} to \eqref{8} by substituting variable redshift function $\phi(r)=\frac{1}{r}$

\begin{eqnarray}
\rho&=&\frac{1}{4 (r-1)^3 r^2}\Bigg[e^{-r-{r_0}} \left(R_c \mu 2^{p+1} p e^{2 r} r \left(p^2 \left(r^2-2\right)^2 r-p \left(2 r^5+2 r^4-12 r^3+r^2+12 r-2\right)\right.\right.\nonumber\\
&+&\left.\left.r^5+2 r^4-8 r^3+r^2+8 r-2\right) \left(\frac{(r-1) e^{{r_0}-r}}{R_c r^2}\right)^p+R_c \mu 2^p r \left(-2 p^3 r \left(r^2-2\right)^2+p^2 \left(5 r^5\right.\right.\right.\nonumber\\
&+&\left.\left.\left.3 r^4-26 r^3+4 r^2+24 r-4\right)+p \left(-3 r^5+r^4+6 r^3+8 r^2-20 r+4\right)-2 (r-1)^3 r\right)\right.\nonumber\\
&\times&\left.e^{r+{r_0}} \left(\frac{(r-1) e^{{r_0}-r}}{R_c r^2}\right)^p-4 (r-1)^4 e^{2 {r_0}}\right)\Bigg]
\end{eqnarray}

\begin{eqnarray}
p_r&=&\frac{1}{2 r^3}\Bigg[\frac{1}{(r-1)^2}\Big[R_c \mu \left(-2^p\right) \left(2 p^2-p-1\right)r^5 \left(\frac{(r-1) e^{{r_0}-r}}{R_c r^2}\right)^p\nonumber\\
&+& R_c \mu 2^p \left(4 p^2-8 p+1\right) r^3 \left(\frac{(r-1) e^{{r_0}-r}}{R_c r^2}\right)^p+2 r^2 \left(R_c \mu 2^p p \right.\nonumber\\
&\times&\left.\left(\frac{(r-1) e^{{r_0}-r}}{R_c r^2}\right)^p-2\right)+R_c \mu 2^p (3 p-2) r^4 \left(\frac{(r-1) e^{{r_0}-r}}{R_c r^2}\right)^p+8 r-4\Big]+\frac{1}{r-1}\nonumber\\
&\times&\Big[\mu 2^{p+1} p \left((p-1) r^3+(3-2 p) r-1\right) \left(\frac{(r-1) e^{{r_0}-r}}{R_c r^2}\right)^{p-1}\Big]-2 (r-2) e^{{r_0}-r}\Bigg]
\end{eqnarray}

\begin{eqnarray}
p_t&=&\frac{1}{4 (1-r)^3 r^4}\Bigg[-R_c \mu 2^p p \left(2 p^2-5 p+3\right) r^8 \left(\frac{(r-1) e^{r_0-r}}{R_c r^2}\right)^p+R_c \mu 2^p \left(p^2+2 p-2\right) r^7 \nonumber\\
&\times&\left(\frac{(r-1) e^{r_0-r}}{R_c r^2}\right)^p+R_c \mu 2^p \left(6 p^2+p-6\right) r^5 \left(\frac{(r-1) e^{r_0-r}}{R_c r^2}\right)^p+\mu 2^{p+1} p (r-1) \left(-2 \left(2 p^2-5p\right.\right.\nonumber\\
&+&\left.\left.3\right) r^4+\left(4 p^2-8 p+3\right) r^2+(p-1)^2 r^6-(p-1) r^5+(3-2 p) r^3-r+1\right) \left(\frac{(r-1) e^{r_0-r}}{R_c r^2}\right)^{p-1}\nonumber\\
&+&R_c \mu 2^p \left(8 p^3-22 p^2+5 p+6\right) r^6 \left(\frac{(r-1) e^{r_0-r}}{R_c r^2}\right)^p-r^4 \left(R_c \mu 2^p \left(8 p^3-16 p^2+9 p-2\right)\right.\nonumber\\
&\times&\left. \left(\frac{(r-1) e^{r_0-r}}{R_c r^2}\right)^p+4\right)-R_c \mu 2^{p+1} p r^2 \left(\frac{(r-1) e^{r_0-r}}{R_c r^2}\right)^p+2 r^3 \left(R_c \mu 2^p p \left(\frac{(r-1) e^{r_0-r}}{R_c r^2}\right)^p+4\right)\nonumber\\
&-&2 (r-1)^3 \left(r^3-r^2-2 r-2\right) e^{r_0-r}-8 r+4\Bigg]
\end{eqnarray}
\begin{eqnarray}
\rho + p_r&=&\frac{1}{4 (r-1)^3 r^2}\Bigg[e^{-r-{r_0}} \left(R_c \mu 2^{p+1} p e^{2 r} r \left(p^2 \left(r^2-2\right)^2 r-p \left(2 r^5+2 r^4-12 r^3+r^2+12 r-2\right)\right.\right.\nonumber\\
&+&\left.\left.r^5+2 r^4-8 r^3+r^2+8 r-2\right) \left(\frac{(r-1) e^{{r_0}-r}}{R_c r^2}\right)^p+R_c \mu 2^p r \left(-2 p^3 r \left(r^2-2\right)^2+p^2 \left(5 r^5\right.\right.\right.\nonumber\\
&+&\left.\left.\left.3 r^4-26 r^3+4 r^2+24 r-4\right)+p \left(-3 r^5+r^4+6 r^3+8 r^2-20 r+4\right)-2 (r-1)^3 r\right)\right.\nonumber\\
&\times&\left.e^{r+{r_0}} \left(\frac{(r-1) e^{{r_0}-r}}{R_c r^2}\right)^p-4 (r-1)^4 e^{2 {r_0}}\right)\Bigg] + \frac{1}{2 r^3}\Bigg[\frac{1}{(r-1)^2}\Big[R_c \mu \left(-2^p\right) \left(2 p^2-p-1\right)\nonumber\\
&\times& r^5 \left(\frac{(r-1) e^{{r_0}-r}}{R_c r^2}\right)^p+R_c \mu 2^p \left(4 p^2-8 p+1\right) r^3 \left(\frac{(r-1) e^{{r_0}-r}}{R_c r^2}\right)^p+2 r^2 \left(R_c \mu 2^p p \right.\nonumber\\
&\times&\left.\left(\frac{(r-1) e^{{r_0}-r}}{R_c r^2}\right)^p-2\right)+R_c \mu 2^p (3 p-2) r^4 \left(\frac{(r-1) e^{{r_0}-r}}{R_c r^2}\right)^p+8 r-4\Big]+\frac{1}{r-1}\nonumber\\
&\times&\Big[\mu 2^{p+1} p \left((p-1) r^3+(3-2 p) r-1\right) \left(\frac{(r-1) e^{{r_0}-r}}{R_c r^2}\right)^{p-1}\Big]-2 (r-2) e^{{r_0}-r}\Bigg]
\end{eqnarray}

\begin{eqnarray}
\rho + p_t&=&\frac{1}{4 (r-1) r^4}\Bigg[e^{-r-{r_0}} \left(R_c \mu \left(-2^{p+1}\right) p e^{2 r} r^2 \left((p-1) r^3+(3-2 p) r+1\right) \left(\frac{(r-1) e^{{r_0}-r}}{R_c r^2}\right)^p\right.\nonumber\\
&+&\left.e^{r+{r_0}} \left(2 r^2 \left(R_c \mu 2^p p \left(\frac{(r-1) e^{{r_0}-r}}{R_c r^2}\right)^p+2\right)+R_c \mu 2^p p (2 p-1) r^5 \left(\frac{(r-1) e^{{r_0}-r}}{R_c r^2}\right)^p\right.\right.\nonumber\\
&-&\left.\left.R_c \mu 2^p p r^4 \left(\frac{(r-1) e^{{r_0}-r}}{R_c r^2}\right)^p-R_c \mu 2^{p+1} p (2 p-3) r^3 \left(\frac{(r-1) e^{{r_0}-r}}{R_c r^2}\right)^p-4\right)\right.\nonumber\\
&-&\left.2 \left(r^4-2 r^3+3 r^2-2\right) e^{2 {r_0}}\right)\Bigg]
\end{eqnarray}

\begin{eqnarray}
\rho + p_r + 2p_t&=&\frac{1}{4 (1-r)^3 r^4}\Bigg[-R_c \mu 2^p p \left(2 p^2-5 p+3\right) r^8 \left(\frac{(r-1) e^{{r_0}-r}}{R_c r^2}\right)^p+R_c \mu 2^p \left(3 p^2+p-4\right)\nonumber\\
&\times& r^7 \left(\frac{(r-1) e^{{r_0}-r}}{R_c r^2}\right)^p+\mu 2^{p+1} p (r-1) \left(-2 \left(2 p^2-5 p+3\right) r^4+\left(4 p^2-8 p+6\right) r^2\right.\nonumber\\
&+&\left.(p-1)^2 r^6-2 (p-1) r^5+(p-1) r^3-2 (p+1) r+2\right) \left(\frac{(r-1) e^{{r_0}-r}}{R_c r^2}\right)^{p-1}\nonumber\\
&+&R_c \mu 2^{p+1} \left(4 p^3-11 p^2+6\right) r^6 \left(\frac{(r-1) e^{{r_0}-r}}{R_c r^2}\right)^p-R_c \mu 2^{p+1} \left(4 p^3-8 p^2+9 p-2\right) r^4\nonumber\\
&\times& \left(\frac{(r-1) e^{{r_0}-r}}{R_c r^2}\right)^p-4 r^2 \left(R_c \mu 2^p p \left(\frac{(r-1) e^{{r_0}-r}}{R_c r^2}\right)^p-6\right)+R_c \mu 2^{p+2} (4 p-3) r^5\nonumber\\
&\times& \left(\frac{(r-1) e^{{r_0}-r}}{R_c r^2}\right)^p+4 r^3 \left(R_c \mu 2^p p (p+1) \left(\frac{(r-1) e^{{r_0}-r}}{R_c r^2}\right)^p-2\right)+4 (r-1)^3 \nonumber\\
&\times& \left(r^2+2\right) e^{{r_0}-r}-24 r+8\Bigg]
\end{eqnarray}

\begin{eqnarray}
\rho-|p_r|&=&\frac{1}{4 (r-1)^3 r^2}\Bigg[e^{-r-{r_0}} \left(R_c \mu 2^{p+1} p e^{2 r} r \left(p^2 \left(r^2-2\right)^2 r-p \left(2 r^5+2 r^4-12 r^3+r^2+12 r-2\right)\right.\right.\nonumber\\
&+&\left.\left.r^5+2 r^4-8 r^3+r^2+8 r-2\right) \left(\frac{(r-1) e^{{r_0}-r}}{R_c r^2}\right)^p+R_c \mu 2^p r \left(-2 p^3 r \left(r^2-2\right)^2+p^2 \left(5 r^5\right.\right.\right.\nonumber\\
&+&\left.\left.\left.3 r^4-26 r^3+4 r^2+24 r-4\right)+p \left(-3 r^5+r^4+6 r^3+8 r^2-20 r+4\right)-2 (r-1)^3 r\right)\right.\nonumber\\
&\times&\left.e^{r+{r_0}} \left(\frac{(r-1) e^{{r_0}-r}}{R_c r^2}\right)^p-4 (r-1)^4 e^{2 {r_0}}\right)\Bigg]-\Bigg|\frac{1}{2 r^3}\Bigg[\frac{1}{(r-1)^2}\Big[R_c \mu \left(-2^p\right) \left(2 p^2-p-1\right)r^5\nonumber\\
&\times& \left(\frac{(r-1) e^{{r_0}-r}}{R_c r^2}\right)^p+ R_c \mu 2^p \left(4 p^2-8 p+1\right) r^3 \left(\frac{(r-1) e^{{r_0}-r}}{R_c r^2}\right)^p+2 r^2 \left(R_c \mu 2^p p \right.\nonumber\\
&\times&\left.\left(\frac{(r-1) e^{{r_0}-r}}{R_c r^2}\right)^p-2\right)+R_c \mu 2^p (3 p-2) r^4 \left(\frac{(r-1) e^{{r_0}-r}}{R_c r^2}\right)^p+8 r-4\Big]+\frac{1}{r-1}\nonumber\\
&\times&\Big[\mu 2^{p+1} p \left((p-1) r^3+(3-2 p) r-1\right) \left(\frac{(r-1) e^{{r_0}-r}}{R_c r^2}\right)^{p-1}\Big]-2 (r-2) e^{{r_0}-r}\Bigg]\Bigg|
\end{eqnarray}

\begin{eqnarray}
\rho-|p_t|&=&\frac{1}{4 (r-1)^3 r^2}\Bigg[e^{-r-{r_0}} \left(R_c \mu 2^{p+1} p e^{2 r} r \left(p^2 \left(r^2-2\right)^2 r-p \left(2 r^5+2 r^4-12 r^3+r^2+12 r-2\right)\right.\right.\nonumber\\
&+&\left.\left.r^5+2 r^4-8 r^3+r^2+8 r-2\right) \left(\frac{(r-1) e^{{r_0}-r}}{R_c r^2}\right)^p+R_c \mu 2^p r \left(-2 p^3 r \left(r^2-2\right)^2+p^2 \left(5 r^5\right.\right.\right.\nonumber\\
&+&\left.\left.\left.3 r^4-26 r^3+4 r^2+24 r-4\right)+p \left(-3 r^5+r^4+6 r^3+8 r^2-20 r+4\right)-2 (r-1)^3 r\right)\right.\nonumber\\
&\times&\left.e^{r+{r_0}} \left(\frac{(r-1) e^{{r_0}-r}}{R_c r^2}\right)^p-4 (r-1)^4 e^{2 {r_0}}\right)\Bigg]-\Bigg| \frac{1}{4 (1-r)^3 r^4}\Bigg[-R_c \mu 2^p p \left(2 p^2-5 p+3\right) r^8\nonumber\\
&\times&\left(\frac{(r-1) e^{r_0-r}}{R_c r^2}\right)^p+R_c \mu 2^p \left(p^2+2 p-2\right) r^7
\left(\frac{(r-1) e^{r_0-r}}{R_c r^2}\right)^p+R_c \mu 2^p \left(6 p^2+p-6\right) r^5\nonumber\\  &\times&\left(\frac{(r-1) e^{r_0-r}}{R_c r^2}\right)^p+\mu 2^{p+1} p (r-1) \left(-2 \left(2 p^2-5p+3\right) r^4+\left(4 p^2-8 p+3\right) r^2\right.\nonumber\\
&+&\left.(p-1)^2 r^6-(p-1) r^5+(3-2 p) r^3-r+1\right) \left(\frac{(r-1) e^{r_0-r}}{R_c r^2}\right)^{p-1}\nonumber\\
&+&R_c \mu 2^p \left(8 p^3-22 p^2+5 p+6\right) r^6 \left(\frac{(r-1) e^{r_0-r}}{R_c r^2}\right)^p-r^4 \left(R_c \mu 2^p \left(8 p^3-16 p^2+9 p-2\right)\right.\nonumber\\
&\times&\left. \left(\frac{(r-1) e^{r_0-r}}{R_c r^2}\right)^p+4\right)-R_c \mu 2^{p+1} p r^2 \left(\frac{(r-1) e^{r_0-r}}{R_c r^2}\right)^p+2 r^3 \left(R_c \mu 2^p p \left(\frac{(r-1) e^{r_0-r}}{R_c r^2}\right)^p\right.\nonumber\\
&+&\left.4\right)-2 (r-1)^3 \left(r^3-r^2-2 r-2\right) e^{r_0-r}-8 r+4\Bigg] \Bigg|
\end{eqnarray}

\begin{eqnarray}
\frac{p_r}{\rho}&=&\frac{2}{r}\Bigg[\frac{1}{(r-1)^2}\Big[R_c \mu \left(-2^p\right) \left(2 p^2-p-1\right)r^5 \left(\frac{(r-1) e^{{r_0}-r}}{R_c r^2}\right)^p\nonumber\\
&+& R_c \mu 2^p \left(4 p^2-8 p+1\right) r^3 \left(\frac{(r-1) e^{{r_0}-r}}{R_c r^2}\right)^p+2 r^2 \left(R_c \mu 2^p p \right.\nonumber\\
&\times&\left.\left(\frac{(r-1) e^{{r_0}-r}}{R_c r^2}\right)^p-2\right)+R_c \mu 2^p (3 p-2) r^4 \left(\frac{(r-1) e^{{r_0}-r}}{R_c r^2}\right)^p+8 r-4\Big]+\frac{1}{r-1}\nonumber\\
&\times&\Big[\mu 2^{p+1} p \left((p-1) r^3+(3-2 p) r-1\right) \left(\frac{(r-1) e^{{r_0}-r}}{R_c r^2}\right)^{p-1}\Big]-2 (r-2) e^{{r_0}-r}\Bigg]\nonumber\\
&\div& \frac{1}{(r-1)^3 }\Bigg[e^{-r-{r_0}} \left(R_c \mu 2^{p+1} p e^{2 r} r \left(p^2 \left(r^2-2\right)^2 r-p \left(2 r^5+2 r^4-12 r^3+r^2+12 r-2\right)\right.\right.\nonumber\\
&+&\left.\left.r^5+2 r^4-8 r^3+r^2+8 r-2\right) \left(\frac{(r-1) e^{{r_0}-r}}{R_c r^2}\right)^p+R_c \mu 2^p r \left(-2 p^3 r \left(r^2-2\right)^2+p^2 \left(5 r^5\right.\right.\right.\nonumber\\
&+&\left.\left.\left.3 r^4-26 r^3+4 r^2+24 r-4\right)+p \left(-3 r^5+r^4+6 r^3+8 r^2-20 r+4\right)-2 (r-1)^3 r\right)\right.\nonumber\\
&\times&\left.e^{r+{r_0}} \left(\frac{(r-1) e^{{r_0}-r}}{R_c r^2}\right)^p-4 (r-1)^4 e^{2 {r_0}}\right)\Bigg]
\end{eqnarray}

\begin{eqnarray}
p_t-p_r&=&\frac{1}{4 r^4}\Bigg[-\frac{1}{(r-1)^2}\Bigg(\mu 2^{p+1} p \left(\left(4 p^2-4 p-5\right) r^2+(p-1)^2 r^6+(p-1) r^5-4 (p-1)^2 r^4\right.\nonumber\\
&+&\left.(9-6 p) r^3+r+1\right) \left(\frac{(r-1) e^{r_0-r}}{R_c r^2}\right)^{p-1}\Bigg)+\frac{1}{(r-1)^3}+2 \left(r^3+r^2-6 r-2\right) e^{r_0-r}\nonumber\\
&\times&\Bigg(R_c \mu 2^p p \left(2 p^2-5 p+3\right) r^8 \left(\frac{(r-1) e^{r_0-r}}{R_c r^2}\right)^p-R_c \mu 2^p p \left(8 p^2-18 p+9\right) r^6 \nonumber\\
&\times&\left(\frac{(r-1) e^{r_0-r}}{R_c r^2}\right)^p+r^4 \left(R_c \mu 2^p p \left(8 p^2-8 p-11\right) \left(\frac{(r-1) e^{r_0-r}}{R_c r^2}\right)^p+12\right)+2 r^2\nonumber\\
&\times& \left(R_c \mu 2^p p \left(\frac{(r-1) e^{r_0-r}}{R_c r^2}\right)^p+12\right)+R_c \mu 2^p p (3 p-4) r^7 \left(\frac{(r-1) e^{r_0-r}}{R_c r^2}\right)^p-7 R_c \mu 2^p p\nonumber\\
&\times& (2 p-3) r^5 \left(\frac{(r-1) e^{r_0-r}}{R_c r^2}\right)^p+2 r^3 \left(R_c \mu 2^p p \left(\frac{(r-1) e^{r_0-r}}{R_c r^2}\right)^p-16\right)-4\Bigg)\Bigg]
\end{eqnarray}
\begin{figure}
	\centering
	\subfigure[$\rho$]{\includegraphics[scale=.37]{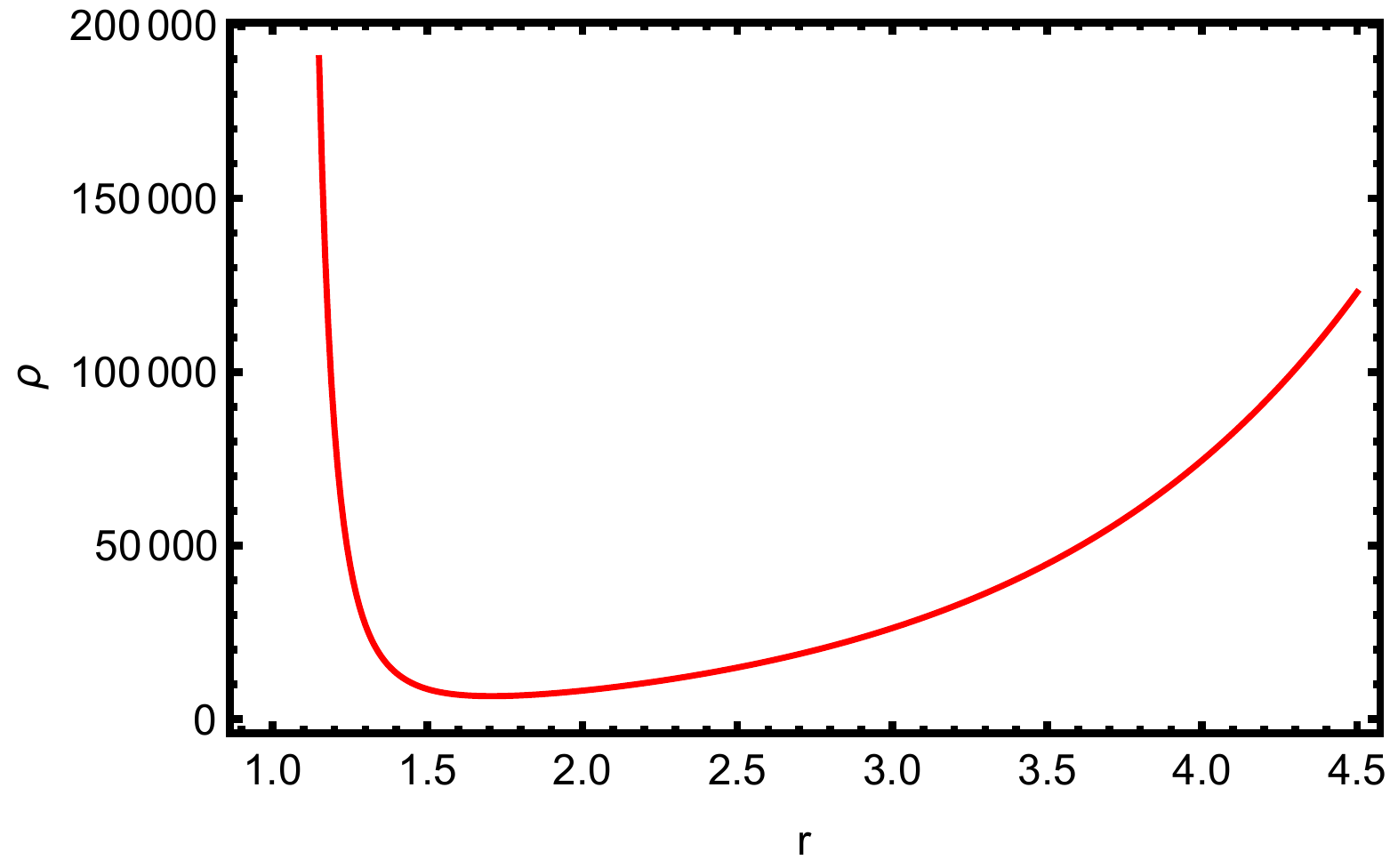}}\hspace{.1cm}
	\subfigure[$\rho+p_r$]{\includegraphics[scale=.37]{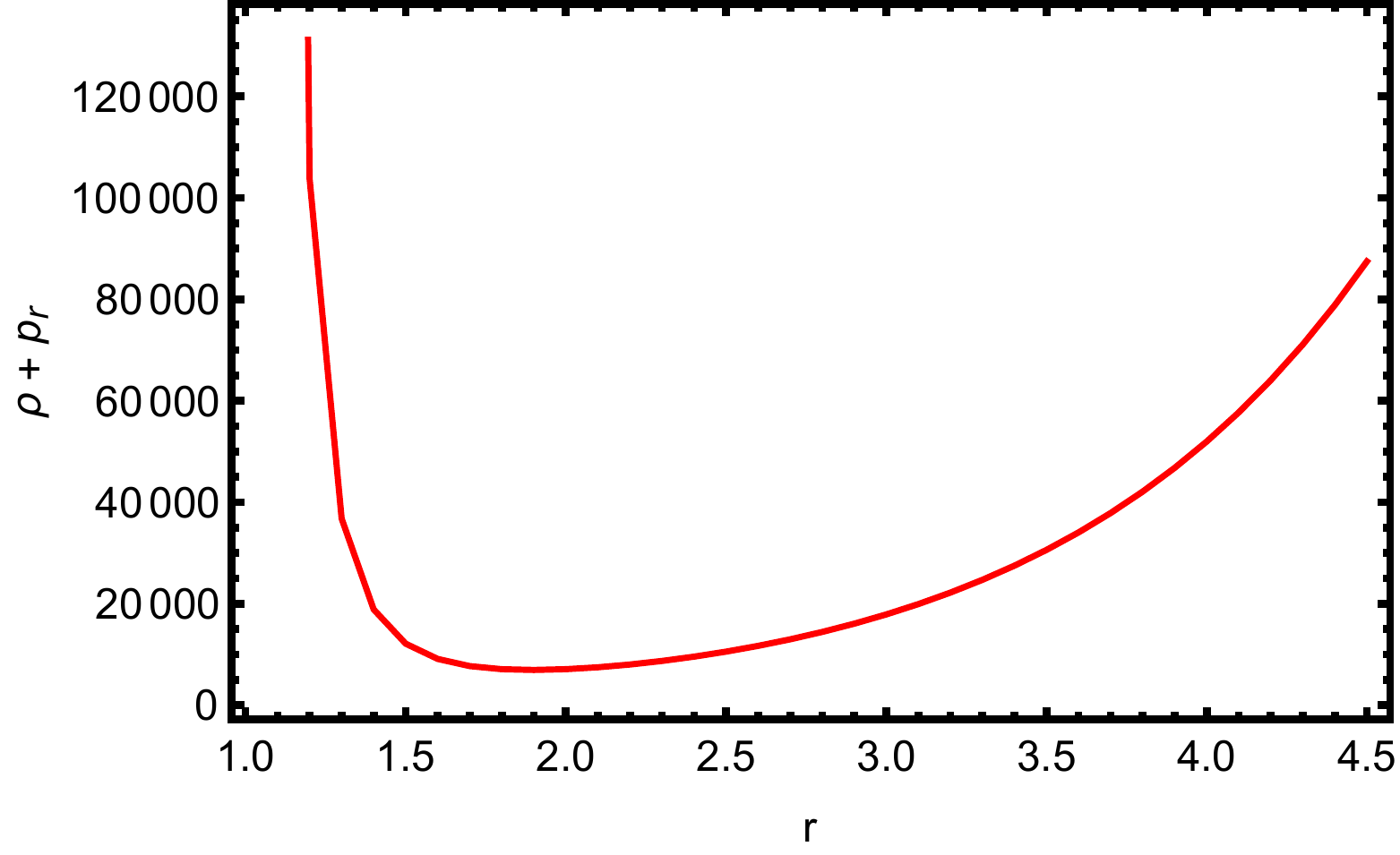}}\hspace{.1cm}
	\subfigure[$\rho+p_t$]{\includegraphics[scale=.37]{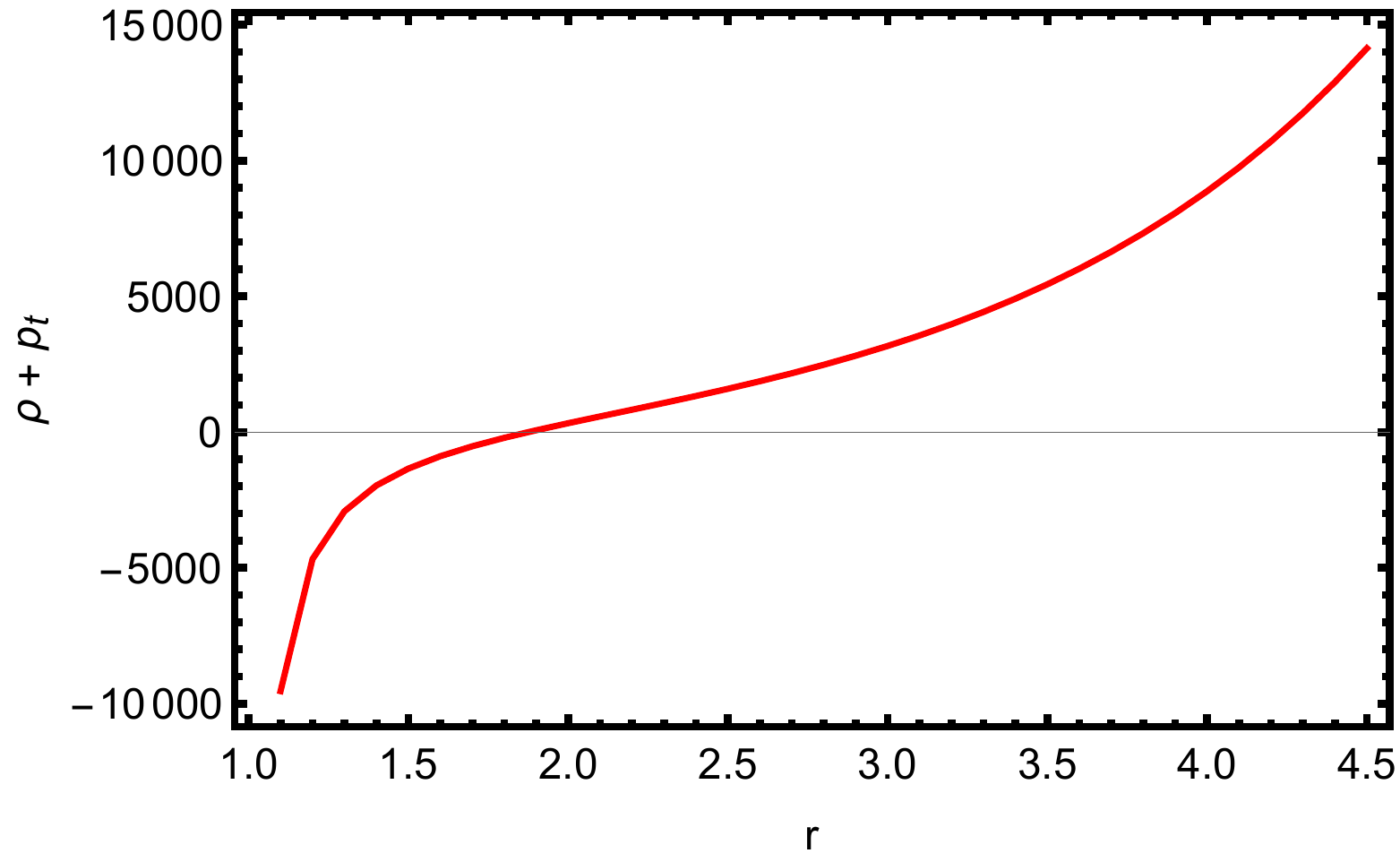}}\hspace{.1cm}
	\subfigure[$\rho-|p_r|$]{\includegraphics[scale=.37]{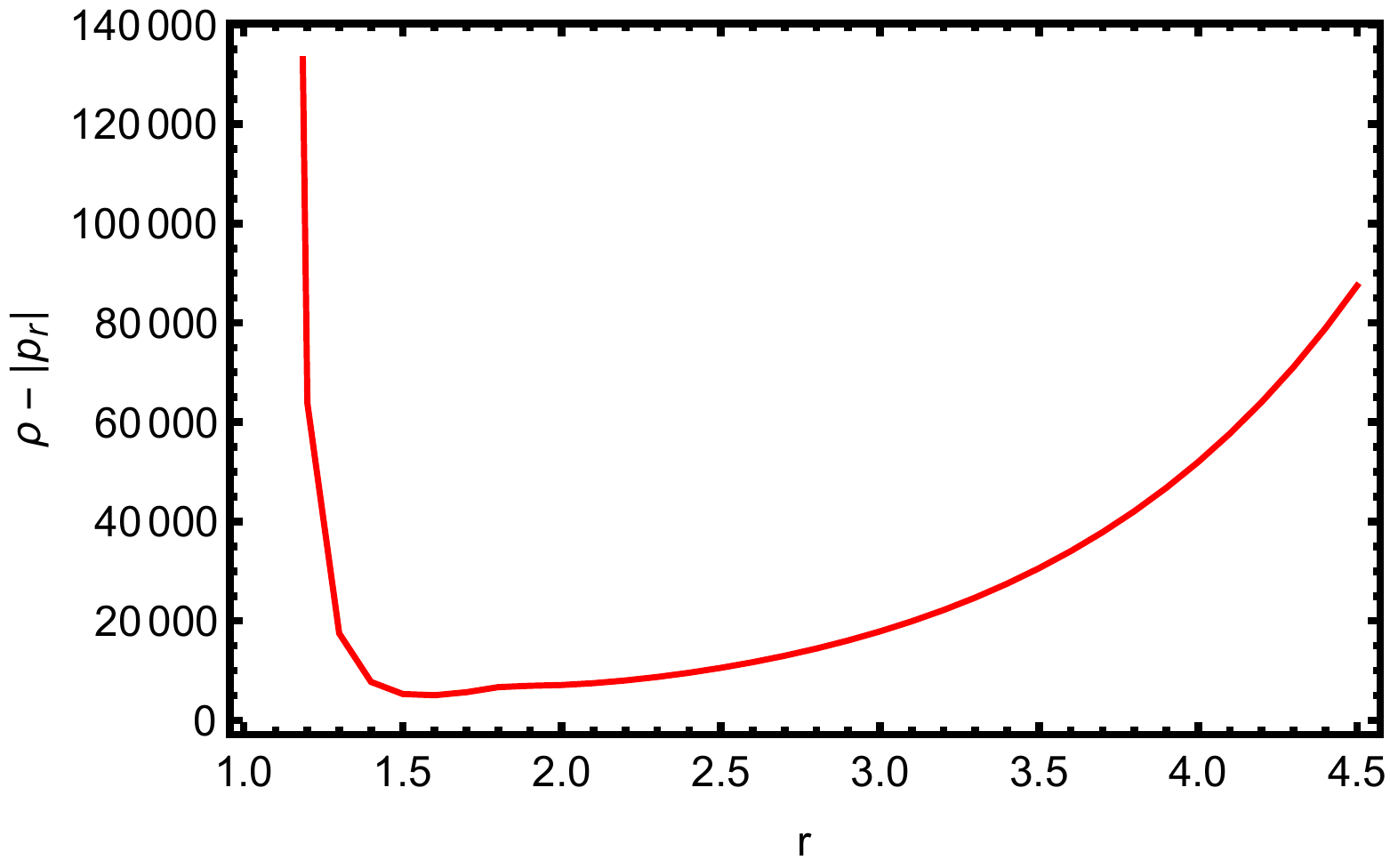}}\hspace{.1cm}
	\subfigure[$\rho-|p_t|$]{\includegraphics[scale=.37]{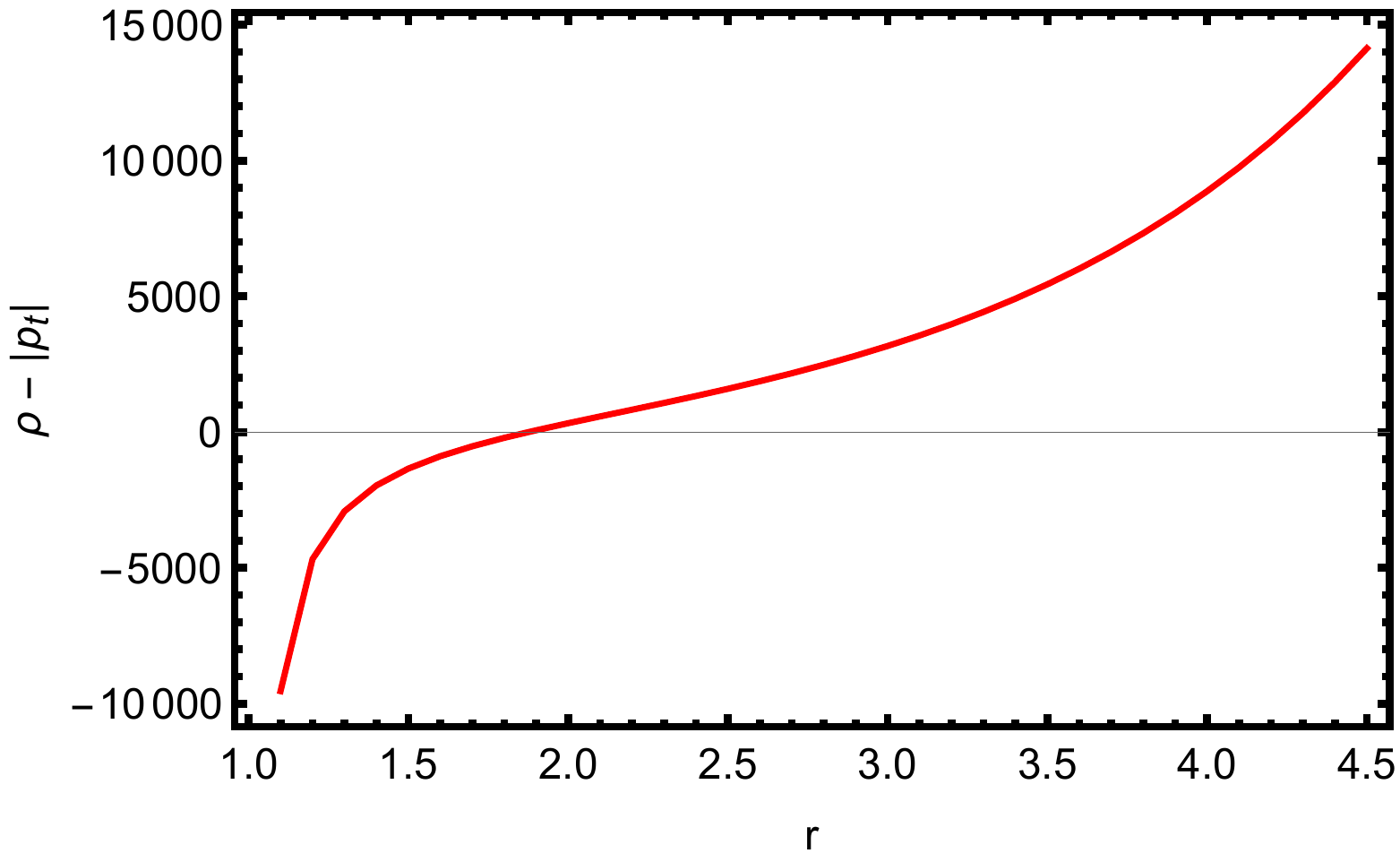}}\hspace{.1cm}
	\subfigure[$\rho+p_r+2p_t$]{\includegraphics[scale=.37]{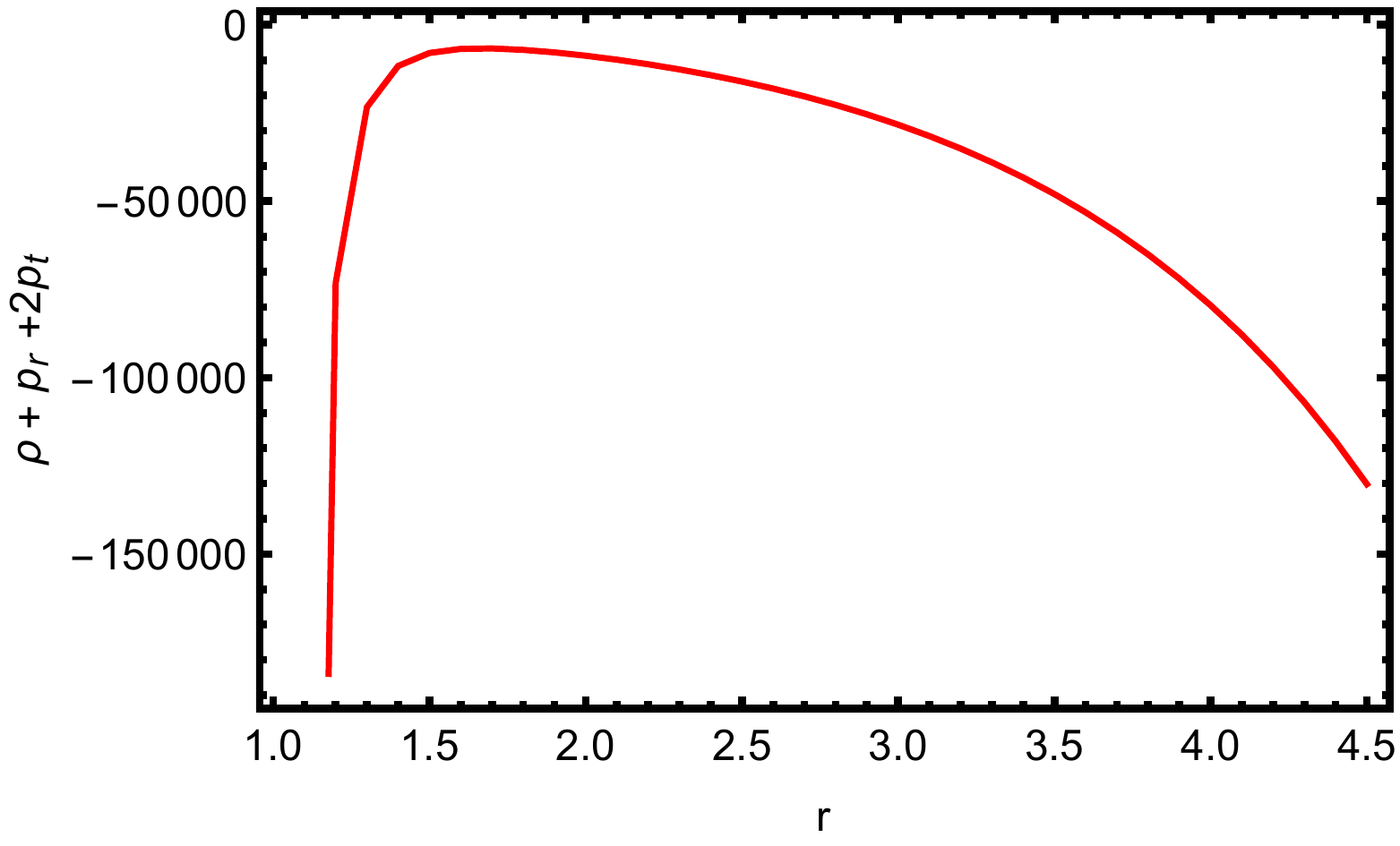}}\hspace{.1cm}
	\subfigure[$\omega$]{\includegraphics[scale=.37]{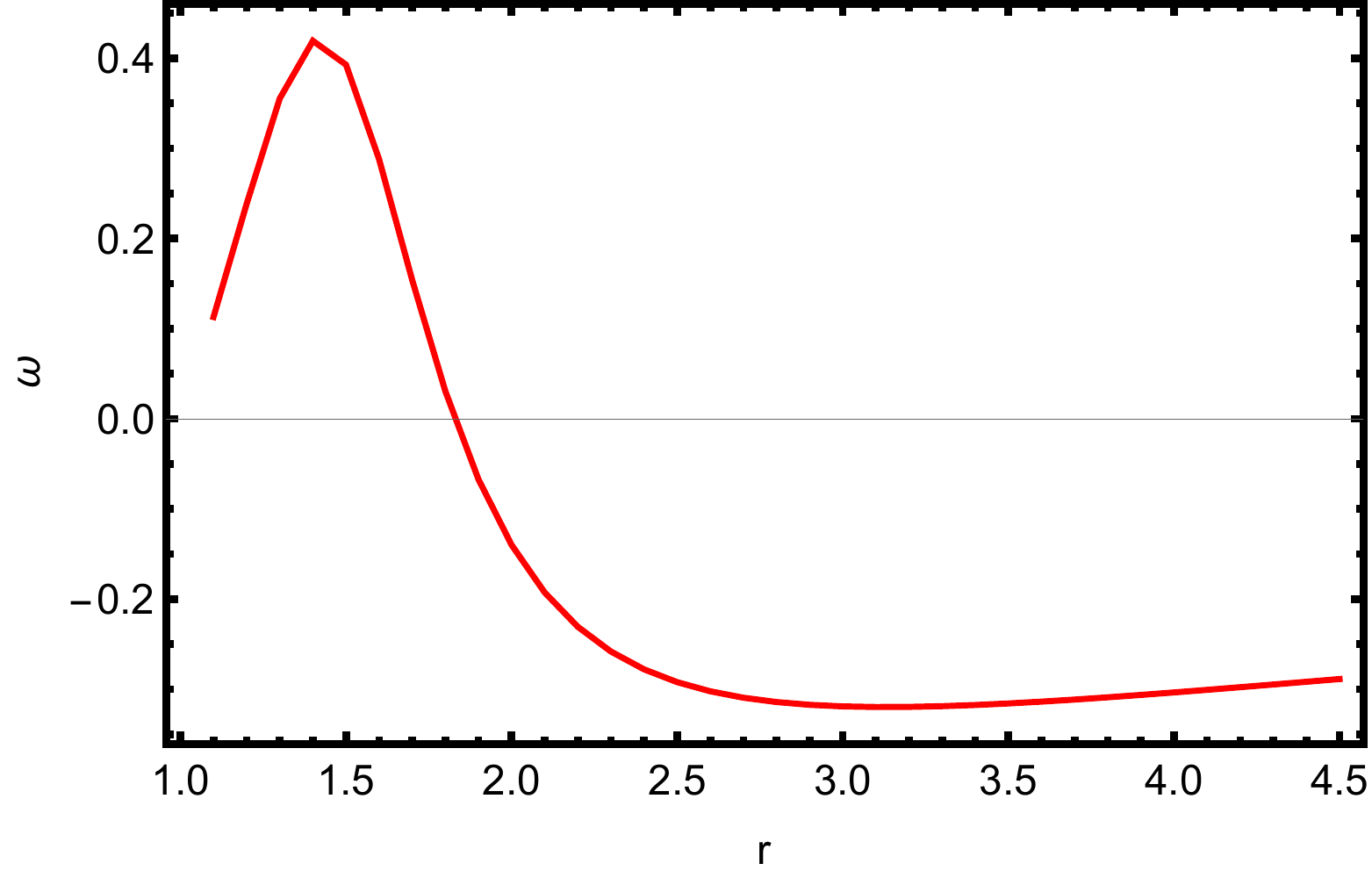}}\hspace{.1cm}
	\subfigure[$\triangle$]{\includegraphics[scale=.37]{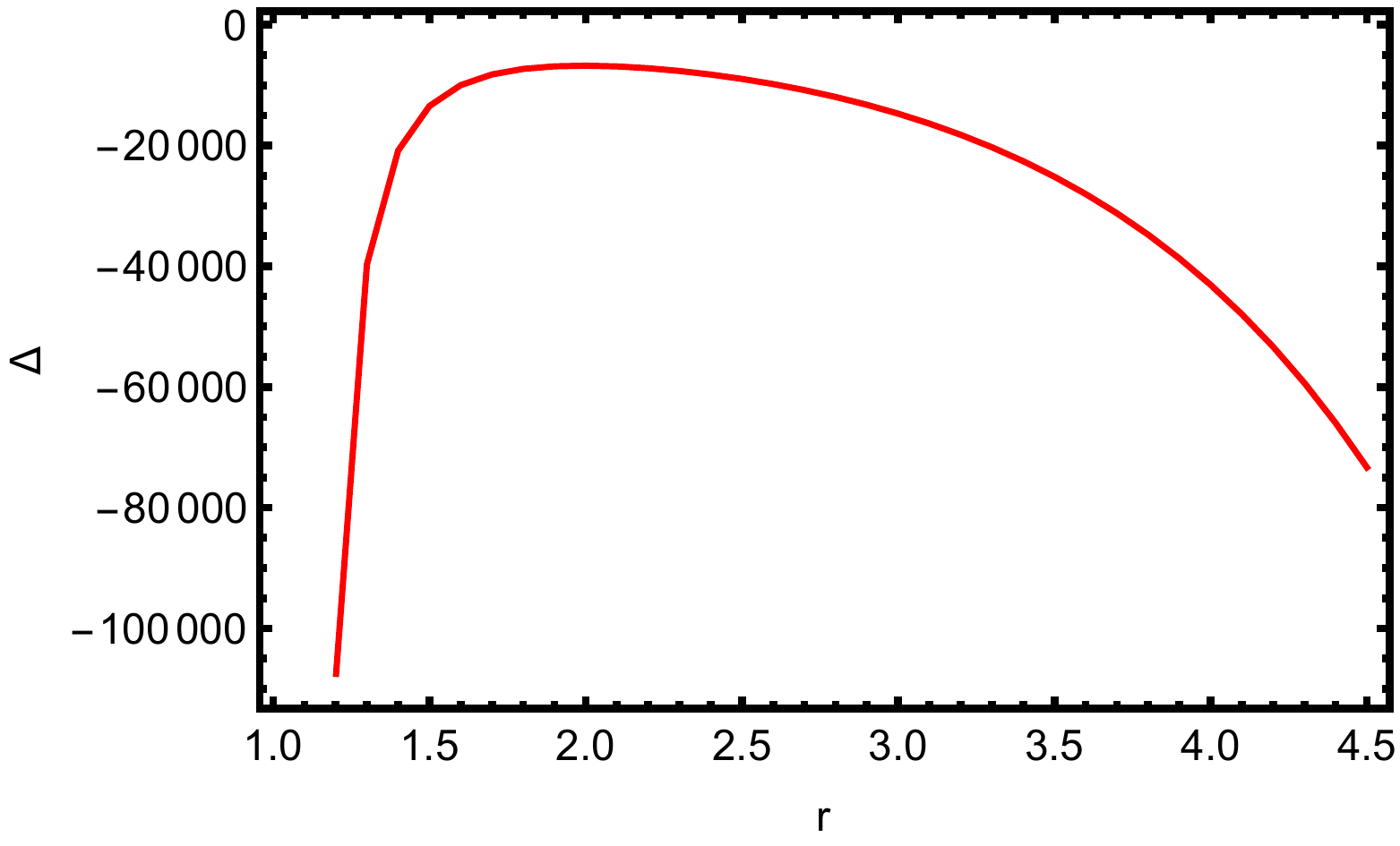}}\hspace{.1cm}
	\caption{Plots for Density, NEC, SEC, DEC, $\triangle$ \& $\omega$ with $\phi(r)=\frac{1}{r}$}
\end{figure}

\section{Results \& Discussion}
In this work, the background of $f(R)$ theory of gravity, a candidate theory explaining the current accelerating scenario of the universe, is taken into account with viable function
$f(R)=R-\mu R_c\Big(\frac{R}{R_c}\Big)^p$, where $\mu$, $R_c>0$ and $0<p<1$, to study traversable wormholes. The wormhole models are described by shape function $b(r)$ and redshift function $\phi(r)$, so these functions play an important role in wormhole modeling. We have considered the shape function $b(r)=\frac{r}{exp(r-r_0)}$ \citep{godani1}. Further, the redshift function  $\phi(r)$ can be a constant or variable. It should   be finite and $e^{\phi(r)}$ should tend to unity as $r$ tends to $\infty$. We have taken $(i)$  $\phi(r)=c$ (constant) and $(ii)$  $\phi(r)=\frac{1}{r}$. Then we have examined the  nature of energy conditions (ECs), namely  null energy condition (NEC),  weak energy condition (WEC), strong energy condition (SEC) and dominant energy condition (DEC), and found the regions where most of the energy conditions are valid. Our model includes two types of pressures: radial pressure $p_r$ and tangential pressure $p_t$. The above  ECs are defined in terms of these pressures in the following way: (i) NEC is said to be satisfied if $\rho + p_r\geq 0$ and $\rho + p_t\geq 0$; (ii) WEC is said to be obeyed if $\rho\geq 0$, $\rho + p_r\geq 0$ and $\rho + p_t\geq 0$; (iii) SEC is said to be validated if $\rho + p_r\geq 0$, $\rho + p_t\geq 0$ and  $\rho + p_r +2p_t\geq 0$; (iv) DEC is said to be fulfilled if $\rho\geq 0$, $\rho - \lvert p_r\rvert \geq 0$ and $\rho - \lvert p_t\rvert \geq 0$. Further, the equation of state in terms of radial pressure is $p_r=\omega \rho$, where $\omega$ is called the equation of state parameter, and the anisotropy parameter $\triangle$ in terms of pressures $p_t$ and $p_r$ is defined as $\triangle=p_t-p_r$ which can be positive, negative or zero. The positive value of $\triangle$ indicates the repulsive nature of the geometry, negative value suggests the attractive nature of the geometry and zero value tells that the geometry is isotropic. We have checked the validity of energy conditions for both type of redshift functions in $f(R)$ gravity as well as general relativity and determined the radius of the throat so that the existence of exotic matter can be neglected and we can get wormhole solutions filled with non-exotic matter. The results in $f(R)$ gravity are discussed in the following two cases for entire range of validity of parameters $\mu$, $p$ and $R_c$: Case 1:  $\phi(r)=c$, where $c$ is a constant and Case 2: $\phi(r)=\frac{1}{r}$.  \\

\noindent
\textbf{Case 1:}  $\phi(r)=c$, where $c$ is a constant\\
In this case, the energy density $\rho$ is found to be a positive function for $r\geq 1.4$, else it has indeterminate or non-real value. This indicates that the exotic matter may be presented near the throat of wormhole only. To avoid this complexity, we can take size of the wormhole throat greater than 1.4 . Then we have checked the nature of null energy condition terms. The first NEC term $\rho+p_r$ is found  to be positive for $r\geq 1.8$ and second NEC term $\rho+p_t$ is obtained to be positive for $r>2.4$. Otherwise, these terms have either negative, indeterminate or non-real values. Hence, the NEC is satisfied for $r>2.4$.  This shows the validity of WEC and hence NEC for $r>2.4$. Further, we have examined SEC and DEC. It is observed that SEC term $\rho + p_r +2p_t$ is negative for $r>1.8$, first DEC term $\rho-|p_r|$ is positive for $r\geq1.8$ and second DEC term $\rho-|p_t|$  is positive for $r>2.4$. This indicates the dissatisfaction of SEC everywhere and satisfaction of DEC for $r>2.4$. Thus, NEC, WEC and DEC are fulfilled for $r>2.4$. Now, the question is: which type of geometric configuration is there for $r>2.4$? To answer this question, we have analyzed the nature of equation of state parameter $\omega$ and anisotropy parameter $\triangle$. We have obtained $\omega$ to lie between 0 and 1 for  $1.8\leq r\leq 2.4$ and to lie between -1 and 0 for  $r>2.4$. Further, $\triangle$ possesses negative values for all $r>1.4$, otherwise it has indeterminate or non-real values. This means the geometry is  attractive and filled with quintessence type matter for every $r>2.4$. Results for this case are also summarized in Table-1.\\

\noindent
\textbf{Case 2:} $\phi(r)=\frac{1}{r}$\\
In this case, the energy density $\rho$ is obtained to be positive for $r\geq 1.2$, otherwise it has indeterminate or non-real values. This depicts the presence of exotic matter near the throat of wormhole like case 1 but for $r\geq 1.2$. Then the first NEC term $\rho+p_r$ is found to be  positive for $r\geq 1.2$ and second NEC term $\rho+p_t$ to be positive for $r>1.8$. Otherwise, these terms have either negative, indeterminate or non-real values. This  shows the validity of both  NEC and WEC for $r>1.8$. Further, we have examined SEC and DEC. It is observed that SEC term $\rho + p_r +2p_t$ is negative for $r>1.2$, first DEC term $\rho-|p_r|$ is positive for $r\geq1.2$ and second DEC term $\rho-|p_t|$  is positive for $r>1.8$. This means that SEC is disobeyed  everywhere and  DEC is obeyed for $r>1.8$. Thus, NEC, WEC and DEC are valid for $r>1.8$. Now, like case 1, again the question is: which type of geometric configuration is there for $r>1.8$? So, we have examined the values taken by $\omega$ and $\triangle$. It is found that $\omega$ lies between 0 and 1 for  $1.2\leq r\leq 1.8$ and  between -1 and 0 for  $r>1.8$. Further, $\triangle$ has negative values for all $r>1.2$, otherwise it has indeterminate or non-real values. Thus, the geometric configuration  is  attractive and filled with ordinary or non-phantom fluid for every $r>1.8$. Results for this case are also summarized in Table-2.

\noindent
Thus, for both forms of redshift function, we have attractive geometric configuration of wormhole solutions in $f(R)$ gravity which are  filled with non-exotic matter and satisfy NEC, WEC and DEC. The only difference is that ECs are obeyed for $r>2.4$ in case 1 while these are obeyed for $r>1.8$ in case 2. In case 1, we can consider the radius of the throat equal to 2.4 and in case 2, it can be taken as 1.8 to get the wormhole solutions filled with non-exotic matter. Since the results are little better in Case 2, that is why we have plotted EC terms, $\omega$ and $\triangle$ only for Case 2 in Fig. (1).  \\

\noindent
Further, it is natural to compare the results in $f(R)$ gravity with the corresponding results in GR. In GR, i.e. for $f(R)=R$, we have detected the nature of energy density and energy conditions.
It is known that NEC is violated in GR. However, there may be a possibility for the presence of non-exotic matter for some range of $r$ because of chosen shape or redshift functions. Using
$b(r)=\frac{r}{exp(r-r_0)}$, we have done investigation of EC terms with respect to both constant redshift function and variable redshift function $\phi(r)=\frac{1}{r}$. For constant redshift function,  the energy density is found to be negative for every value of $r$. For  $\phi(r)=\frac{1}{r}$,   the energy density is positive only for $r\in(0,1)$ but the first NEC term is not positive for any value of $r$. This implies the violation of NEC for every value of $r$,  in case of GR. This study suggests that the exotic matter is required to obtain a wormhole solutions in general relativity. Hence, we have obtained the existence of wormhole solutions without support of exotic matter with attractive geometry in viable $f(R)$ gravity with the radius of throat $r_0=2.4$ (for constant $\Phi$) or $r_0=1.8$ (for variable $\Phi$).

\begin{table}[!h]
	\centering
	\caption{Results for  $f(R)=R-\mu R_c\Big(\frac{R}{R_c}\Big)^p$ with $\phi(r)=c$}
	\begin{tabular}{|c|c|l|}
		\hline
		S.No.& Terms& Results\\
		\hline
		1 & $\rho$ & $>0$, for $r\geq 1.4$\\
		&        & indeterminate or imaginary, otherwise\\
		\hline
		2 & $\rho+p_r$ & $>0$, for $r\geq 1.8$\\
		&        & indeterminate or imaginary, otherwise\\
		\cline{1-3}
		3 & $\rho+p_t$ & $>0$, for $r>2.4$\\
		&        & $<0$, for $r\in (1,2.4]$\\
		&        & indeterminate or imaginary, otherwise\\
		\cline{1-3}
		4 & $\rho+p_r+2p_t$ & $<0$, for $r>1.8$\\
		&        & indeterminate or imaginary, otherwise\\
		\cline{1-3}
		5 & $\rho-|p_r|$ & $>0$, for  $r\geq 1.8$\\
		&        & indeterminate or imaginary, otherwise\\
		\cline{1-3}
		6 & $\rho-|p_t|$ & $>0$, for $r>2.4$\\
		&        & $<0$, for $r\in (1,2.4]$\\
		&        & indeterminate or imaginary, otherwise\\
		\cline{1-3}	
		7 & $\omega$ & Between 0 and 1, for $1.8\leq r\leq 2.4$\\
		&        & Between -1 and 0, for $r>2.4$\\
		&        & indeterminate or imaginary, otherwise\\
		\cline{1-3}	 	
		8 & $\triangle$ & $<0$, for $r>1.4$\\
		&        & indeterminate or imaginary, otherwise\\
		\cline{1-3}
	\end{tabular}
\end{table}

\begin{table}[!h]
	\centering
	\caption{Results for  $f(R)=R-\mu R_c\Big(\frac{R}{R_c}\Big)^p$ with $\phi(r)=\frac{1}{r}$}
	\begin{tabular}{|c|c|l|}
		\hline
		S.No.& Terms& Results\\
		\hline
		1 & $\rho$ & $>0$, for $r\geq 1.2$\\
		&        & indeterminate or imaginary, otherwise\\
		\hline
		2 & $\rho+p_r$ & $>0$, for $r\geq 1.2$\\
		&        & indeterminate or imaginary, otherwise\\
		\cline{1-3}
		3 & $\rho+p_t$ & $>0$, for $r>1.8$\\
		&        & $<0$, for $r\in (1,1.8]$\\
		&        & indeterminate or imaginary, otherwise\\
		\cline{1-3}
		4 & $\rho+p_r+2p_t$ & $<0$, for $r>1.2$\\
		&        & indeterminate or imaginary, otherwise\\
		\cline{1-3}
		5 & $\rho-|p_r|$ & $>0$, for  $r\geq 1.2$\\
		&        & indeterminate or imaginary, otherwise\\
		\cline{1-3}
		6 & $\rho-|p_t|$ & $>0$, for $r>1.8$\\
		&        & $<0$, for $r\in (1,1.8]$\\
		&        & indeterminate or imaginary, otherwise\\
		\cline{1-3}	
		7 & $\omega$ & Between 0 and 1, for $1.2\geq r\geq 1.8$\\
		&        & Between -1 and 0, for $r>1.8$\\
		&        & indeterminate or imaginary, otherwise\\
		\cline{1-3}	 	
		8 & $\triangle$ & $<0$, for $r>1.2$\\
		&        & indeterminate or imaginary, otherwise\\
		\cline{1-3}
	\end{tabular}
\end{table}

\section{Conclusion}
Wormholes have been introduced by Morris and Thorne \citep{morris1} as a medium for teaching general relativity and are detected in many aspects in literature. These are studied in generalized theories of gravity in order to get wormhole objects filled with the matter that obeys the energy conditions. In this article, we have determined wormhole solutions using the  $f(R)$ theory of gravity with viable cosmological model $f(R)=R-\mu R_c\Big(\frac{R}{R_c}\Big)^p$, where $\mu$, $R_c>0$ and $0<p<1$. We have considered constant as well as variable redshift functions with shape function $b(r)=\frac{r}{exp(r-r_0)}$ to solve the field equations for wormholes in $f(R)$ gravity. Further, the energy conditions and geometric configuration are detected. It is found that SEC is violated everywhere with both forms of redshift function. However, NEC, WEC and DEC are valid in both cases for a wide range of radial coordinate $r$. It is concluded that for wormholes having radius of throat $> 1.8$ with variable redshift function or $> 2.4$ with constant redshift function, exotic type matter is absent completely, i.e. we have wormhole geometries free from exotic matter. Furthermore, we did not found wormhole solutions in GR without exotic matter for any value of $r$. Thus, there is a large difference between the results of two theories. Hence, the $f(R)$ gravity with the model undertaken strongly supports the existence of wormhole solutions filled with the non-exotic matter respecting the energy conditions.\\

\noindent
{\bf Acknowledgment:} The authors are very much thankful to the reviewer and editor for their constructive comments for the improvement of the paper.


\end{document}